\documentclass
[preprint,10pt,twocolumn,prl,letterpaper,noshowpacs,nobalancelastpage]{revtex4}%
\usepackage{amsfonts}
\usepackage{amsmath}
\usepackage{amssymb}
\usepackage[dvips]{graphicx}
\usepackage{hyphenat}%
\providecommand{\U}[1]{\protect\rule{.1in}{.1in}}

\begin{document}
\preprint{ }
\title{Suppressing quantum effects by optically-driven nonequilibrium phonons}
\author{Z. Ovadyahu}
\affiliation{Racah Institute of Physics, The Hebrew University, Jerusalem 9190401, Israel }


\begin{abstract}
Optically-generated nonequilibrium phonon-distribution is used for exploring
the origin of a nonlocal adiabatic response in an interacting Anderson
insulator. Exposing the system to weak infrared radiation is shown to
effectively suppress a long-range effect observed in field-effect experiments
while producing little heating and barely changing the system conductance.
These effects are shown to be consistent with the quantum nature of the effect
and therefore are peculiar to disordered systems that are quantum-coherent.

\end{abstract}
\maketitle

The wavefunctions of Anderson insulators are commonly visioned as decaying
exponentially in space away from their localization center. Localized
wavefunctions with more complicated envelopes, extending over scales longer
than the localization length, $\xi$ are exponentially rare. Associated with
this view is the premise that correlations between spatially-separated states
may be neglected. In case of static disorder, this is a plausible approach for
most realistic situations. When the system is exposed to a time-dependent
potentials however, correlations between wavefunctions of distant sites may
appear that could significantly modify transport properties of the system.

In this spirit, Khemani et al \cite{1} proposed that slowly varying the
potential applied to a site localized at $\vec{r}$ would produce a disturbance
in other localized-sites that are many $\xi$'s apart from $\vec{r}$. The local
potential-change $\partial$V must however be slow enough such that during
$\partial$t the process remains adiabatic. This condition leads to a spatial
range influenced by the local potential-manipulation, r$_{\text{inf}}$,
\cite{1}:%
\begin{equation}
\text{r}_{\text{inf}}\approx\xi\text{\textperiodcentered}\ln\left(
\frac{\text{W}^{\text{2}}}{\partial\text{V/}\partial\text{t\textperiodcentered
}\hbar}\right)
\end{equation}
where W is the disorder responsible for the localization, $\partial
$V/$\partial$t is the rate of the local potential change. In the presence of a
finite phase-coherent time $\tau_{\phi}$, the range is limited to L$_{\phi}$
$\approx\xi$\textperiodcentered$\ln$(W$\tau_{\phi}$/$\hbar$) \cite{1} or by
r$_{\text{inf}}$, whichever is smaller. Accordingly, the cutoff length at high
temperature (or slow $\partial$V/$\partial$t) would be set by L$_{\phi}$,
which due to lack of theory, has to be estimated on the basis of independent
measurement. Note that L$_{\phi}$ depends on the coherence time but not on the
diffusion constant, unlike the situation in weak-localization. Also, the
prefactor in both Eq.1 and L$_{\phi}$ are yet to be determined by a proper
theoretical model.

This quantum-mechanical effect was recently considered as a possible
explanation for a long-range influence of varying the local potential in
field-effect measurements performed on electron-glasses \cite{2}. In the
experiment, the surface potential at the bottom of a sample, controlled by a
gate, affected its entire thickness of 82nm, $\approx$20 times larger than the
localization length.

A crucial aspect of this nonlocal mechanism, heretofore untested, is that the
rate of potential-change, $\partial$V/$\partial$t is a dynamic parameter that
controls the range of influence. Sweeping the gate-voltage V$_{\text{g}}$ as a
means of manipulating $\partial$V/$\partial$t is a viable technique to keep
the process adiabatic but being limited to relatively small frequencies it is
not an effective way to test the logarithmic dependence (Eq.1); To reduce the
logarithmic factor that is typically $\approx$20-30 by a factor of two,
$\partial$V/$\partial$t has to be increased by 6-8 orders of magnitude. This
range of frequencies cannot be implemented with a standard field-effect
technique. As an alternative we employ in this work optically-generated
nonequilibrium-phonons to modulate the sites potential with THz frequencies.
It turns out that even a very weak intensity of this disturbance dramatically
suppresses the long-range effect associated with r$_{\text{inf}}$ while the
system effective temperature and conductance are barely affected. This
technique, which has been used in a number of solid-state projects \cite{3},
is contrasted with another nonequilibrium steady-state (NESS) by measuring the
system conductance away from linear-response. The results demonstrate the
diversity of outcomes obtainable by subjecting a quantum system to different
NESS protocols. In particular, the outcome of a protocol depends on the
frequency of the drive, not on its intensity, in similar vein with the
photoemission versus thermal-emission phenomenon. It has also bearing on
energy-transfer issues of open quantum systems \cite{4}.

Like all quantum effects, the long-range of influence produced by changing the
potential locally is ultimately limited by the phase-coherence of the medium.
The experimental results of this work confirm that coherence length comparable
with what is achievable in diffusive systems at low temperatures are
attainable in some systems. Apparently however, this property is less common
than seems to be generally recognized.%

\begin{figure}[ptb]%
\centering
\includegraphics[
height=4.7435in,
width=3.4411in
]%
{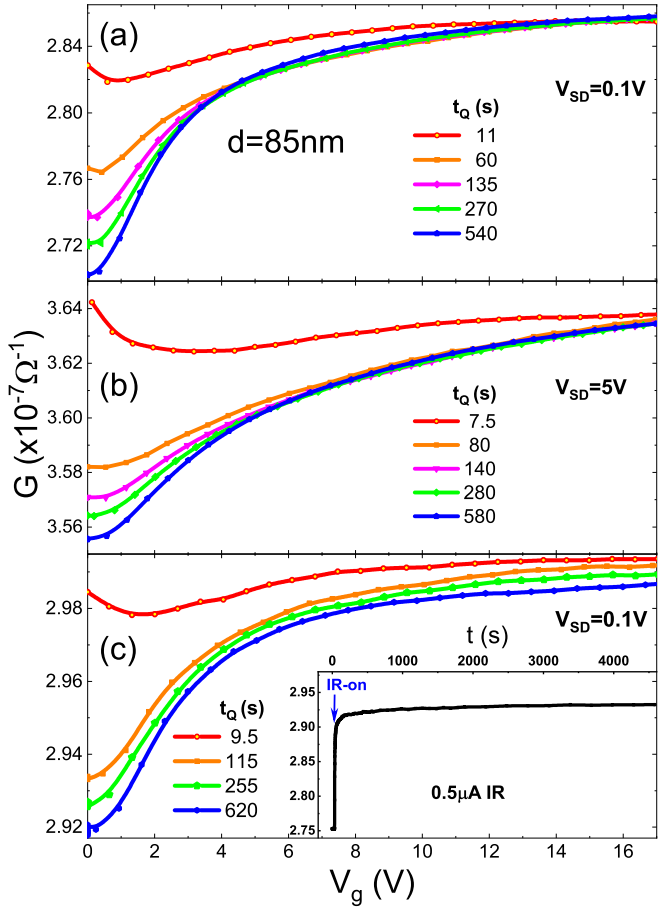}%
\caption{Plates 1a, 1b and 1c show the results of field effect scans taken
under source-drain voltage V$_{\text{SD}}$ indicated for each series of
measurements (see text for description of the protocols). Inset to plate 1c
shows the conductance vs. time after turning on the infrared source operating
with 0.5$\mu$A current and 1.002V forward-bias.}%
\end{figure}

The samples used in this study were amorphous indium-oxide In$_{\text{x}}$O
films with low carrier-concentration \textit{N}$\approx$(1$\pm$%
0.1)x10$^{\text{19}}$cm$^{\text{-3}}$. These were configured as MOSFET devices
for field-effect measurements \cite{5}.

The main results of this study are summarized in the set of measurements
presented in Fig.1. These include four experiments performed on a single
sample with thickness d=85$\pm$20nm and relatively high degree of disorder;
the Ioffe-Regel parameter k$_{\text{F}}\ell$ of this sample is $\approx$0.04,
deep on the insulating side.

Fig.1a, Fig.1b, and Fig.1c show a series of conductance vs. gate-voltage
G(V$_{\text{g}}$) traces taken at different times t$_{\text{Q}}$. The time is
measured from the moment the system was quenched from a random state and
allowed to relax at the measurements temperature T=4.11K. Exciting the system
to achieve randomization is accomplished by exposing the sample to an AlGaAs
diode generating infrared radiation at a wavelength $\approx$0.85$\pm$%
0.05$\mu$m. This device was mounted on the sample-stage 12-15mm from the
sample surface. This is done, using a computer-controlled Keithley 220, by
passing a current of 0.5mA (with forward bias =1.3V) through the diode for 3 seconds.

Shortly after the brief exposure to the IR source is terminated, the
G(V$_{\text{g}}$,t$_{\text{Q}}$) traces are recorded. Each trace is taken from
V$_{\text{g}}$=0 to V$_{\text{g}}$=14V with a constant rate of 0.6V/s. Between
subsequent scans the sample is kept under V$_{\text{g}}$=0 allowing the sample
to relax as manifested by a memory-dip (MD) \cite{6,7} growing deeper with
time. Figure 1b shows the results of the same protocol as in Fig.1a except
that the G(V$_{\text{g}}$,t$_{\text{Q}}$) traces were taken using a non-ohmic
source-drain bias. The non-ohmic measurements resulted in higher conductance,
and wider and shallower MD (see, Fig.2).%
\begin{figure}[ptb]%
\centering
\includegraphics[
height=3.301in,
width=3.4411in
]%
{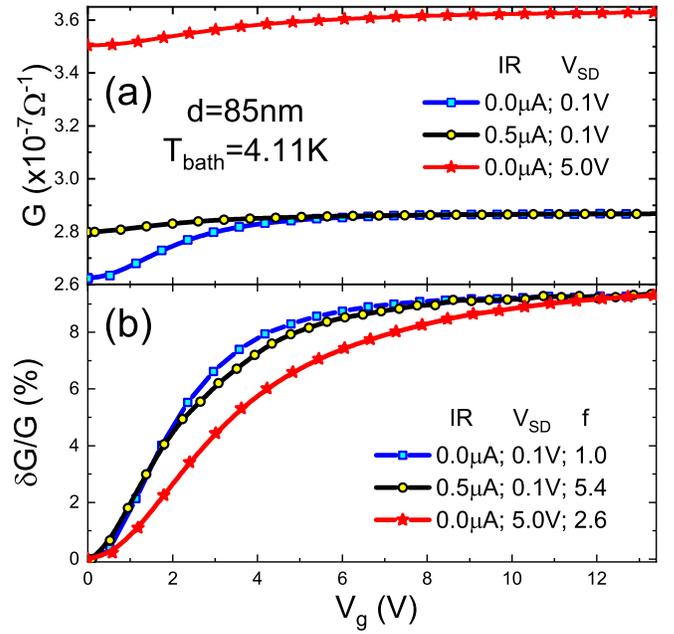}%
\caption{The G(V$_{\text{g}}$) of the 85nm sample taken after several hours of
relaxation at V$_{\text{g}}$=0V. Curves are labeled by the measurement
conditions. (a) The raw G(V$_{\text{g}}$) curves. (b) Comparing the MD shapes
(derived from the raw data) after normalizing them by a constant factor f to
match the linear-response curve at V$_{\text{g}}$=0V and V$_{\text{g}}$=13V.
Note the difference in f between the two NESS protocols.}%
\end{figure}

A modified protocol was used to obtain the data shown in Fig.1c. This protocol
starts by allowing the sample to reach a steady-state under a weak IR
radiation; about one thousandth of the intensity used for completely
randomizing the system. Turning on this radiation promptly increases the
system conductance that, in this sample, eventually settles at a $\approx$6\%
higher than its equilibrium value as shown in Fig.1c. The rest of the protocol
involves the same steps as in Fig.1a, and Fig.1b (the weak IR is kept
\textit{throughout} the duration\textit{ }of this protocol). Note that the
G(V$_{\text{g}}$,t$_{\text{Q}}$) curves in Fig.1c become parallel over the
range of V$_{\text{g}}$ where the G(V$_{\text{g}}$,t$_{\text{Q}}$) traces in
Fig.1a, and Fig.1b are converging to a common plot.

Between protocols the sample was allowed to recover for at least 24 hours.
Each protocol in Fig.1 was repeated twice to ascertain reproducibility.

A simple explanation for the results in Fig.1c is that the charge added by
$\delta$V$_{\text{g}}$ affects only the part of the sample in the vicinity of
the sample-spacer interface while the rest of it is unaffected. In this case
the sample is effectively composed of two conductors in parallel; one where
the G(V$_{\text{g}}$,t$_{\text{Q}}$) curves are like the pattern exhibited by
the sample in Fig.1a, and another for which G(t) just monotonically decreases
after being "quench-cooled", independent of $\delta$V$_{\text{g}}$.
Superimposing these two components reproduces the G(V$_{\text{g}}%
$,t$_{\text{Q}}$) curves exhibited in Fig.1c.

By contrast, using either, the linear-response or the non-ohmic protocols
(Fig.1a and Fig.1b respectively), the converging G(V$_{\text{g}}$%
,t$_{\text{Q}}$) traces suggest that changing the gate-voltage affects the
entire sample volume. This long-range influence, first seen on a thin sample
(5nm In$_{\text{2}}$O$_{\text{3-x}}$ film \cite{6}) is shown here to extend up
to 85nm.

Comparing Fig.1b with Fig.1c and the conditions under which the respective
data were taken, one may conclude the following:

The reason for the suppression of the long range effect by the IR radiation is
\textit{neither due to a change of the system conductance nor due to heating}:
Note that heating and conductance-enhancement are much more conspicuous in the
V$_{\text{SD}}$=5V protocol while leaving the long-range effect essentially
intact (Fig.1b).

Another feature that clearly marks the V$_{\text{SD}}$=5V protocol as the
"hotter" state is shown in Fig.2b. It illustrates how the shape of the MD, for
the two different NESS protocols, deviates from the MD taken under
ohmic-conditions and in the dark. The non-ohmic MD shape is significantly
broader than the ohmic curve and it is consistent with the G(V$_{\text{g}}$)
obtained at elevated temperatures \cite{6,8}. Intriguingly, G(V$_{\text{g}}$)
taken under the weak-IR, while it is reduced in magnitude by a factor of
$\approx$5.4 (as compared with just $\approx$2.6 for the 5V bias, see Fig.2),
its shape deviates from the "reference" much less than in the non-ohmic case.
The latter feature is consistent with the smaller excess energy imparted to
the system which is not surprising given the large disparity in the power
invested in the system by the different sources; The power associated with
V$_{\text{SD}}$=5V is $\approx$20 times larger than that generated by the IR
diode. The more pronounced effect of the IR on the MD magnitude, $\delta$G/G
is a result of the different way the energy absorbed from the 5V bias and IR
fields is distributed among the system degrees of freedom.

If the change of the sample conductance $\Delta$G under either NESS protocol
is related to an excess temperature $\Delta$T by using the G(T) of the sample
\cite{9} then $\Delta$T$\cong$40mK under the weak IR and $\Delta$T$\cong$300mK
for the non-ohmic bias condition. It should be observed that while G(T) is an
equilibrium result $\Delta$G is measured under NESS conditions. Therefore,
these $\Delta$T's should be regarded as merely order of magnitude values and
will be treated as such in our estimates for r$_{\text{inf}}$ detailed below.

We now show that the results of the experiments described above can be
heuristically accounted for by the quantum scenario proposed in \cite{1} and
the peculiarity of the nonequilibrium phonon distribution resulting from the
weak-IR illumination.

The range of disturbance created by a local change of the potential is limited
by either L$_{\phi}$ or by $\partial$V/$\partial$t. Using Eq.1 and L$_{\phi}$
$\approx\xi$\textperiodcentered$\ln$(W$\tau_{\phi}$/$\hbar$) with W$\approx
$0.4eV, $\tau_{\phi}\approx$10$^{\text{-5}}$s, $\xi\approx$5nm, and $\partial
$V/$\partial$t$\approx$0.6meV/s \cite{10} yields r$_{\text{inf}}\approx$
L$_{\phi}\approx$100nm. This agrees with our linear-response results (Fig.1a).
The situation changes when the system is exposed to the IR source. This
initiates a cascade process \cite{11}: Electrons are excited to high-energy,
then relax by phonons emission. The NESS that eventually sets-in (Fig.1c),
sustains vibrations at typical phonon frequencies f$_{\text{ph}}\approx
$10$^{\text{13}}$s$^{\text{-1}}$; the energy of optical-phonons in
In$_{\text{x}}$O is spread over the interval of 350-500cm$^{\text{-1}}$
(0.043-0.062eV) \cite{12},. The associated excess energy sustains a
potential-amplitude $\partial$V$_{\text{IR}}$. The NESS that characterizes the
weak-IR protocol involves a flow of energy from the electronic system to the
phonons and into the bath. Given the small in-going power involved, the lack
of detailed balance may be neglected and it is then plausible to assume that
the electronic system is close to equilibrium with the phonons. Then
$\partial$V$_{\text{IR}}$ may be estimated from:%

\begin{equation}
e\text{\textperiodcentered}\partial\text{V}_{\text{IR}}\approx\text{C}%
_{\text{e}}\text{\textperiodcentered}\Delta\text{T}\approx\frac{\pi^{2}}%
{2}\text{k}_{\text{B}}\left(  \frac{\text{T}}{\text{T}_{\text{F}}}\right)
\Delta\text{T}%
\end{equation}
where C$_{\text{e}}$ is heat-capacity per electron, k$_{\text{B}}$ is the
Boltzmann constant, and T$_{\text{F}}$ is the Fermi temperature. With
T$_{\text{F}}\approx$750K, and $\Delta$T$\cong$40mK noted above, $\partial
$V$_{\text{IR}}$ is of the order of $\approx$0.1$\mu$V giving $\partial
$V$_{\text{IR}}$\textperiodcentered f$_{\text{ph}}$ that is about $\approx
$10$^{\text{9}}$ times larger than the respective value for the
linear-response conditions. Inserted in Eq.1 this factor reduces the
linear-response value r$_{\text{inf}}\approx$100nm to r$_{\text{inf}}\approx
$60nm. This means that, under the weak-IR, the range-of-influence becomes
smaller than the film thickness consistent with the results shown in Fig.1c.%
\begin{figure}[ptb]%
\centering
\includegraphics[
height=2.9352in,
width=3.4411in
]%
{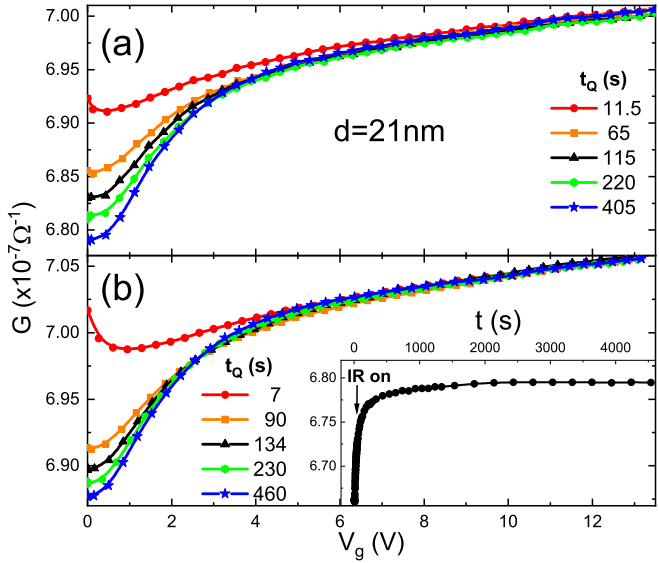}%
\caption{The results of performing G(V$_{\text{g}}$,t$_{\text{Q}}$)
spatial-range test-protocols on a 21nm In$_{\text{x}}$O film with
k$_{\text{F}}\ell$=0.061: (a) The same protocol as used in Fig.1a. (b) The
same conditions as in Fig.1c. The inset in (b) depicts the conductance vs.
time under the 0.5$\mu$A IR used during the G(V$_{\text{g}}$,t$_{\text{Q}}$)
weak-IR protocol.}%
\end{figure}

A natural further test of this conjecture, is to apply the same protocol on
samples that are thinner than 60nm. Three thin ($\approx$20nm) In$_{\text{x}}%
$O batches were prepared for this purpose, with the same composition as the
85nm sample, attempting to match all relevant material parameters. Results of
these measurements are shown in Fig.3 and Fig.4 for two of the samples, one
with resistance smaller and the other with larger resistance than that of the
85nm sample.
\begin{figure}[ptb]%
\centering
\includegraphics[
height=2.9948in,
width=3.4411in
]%
{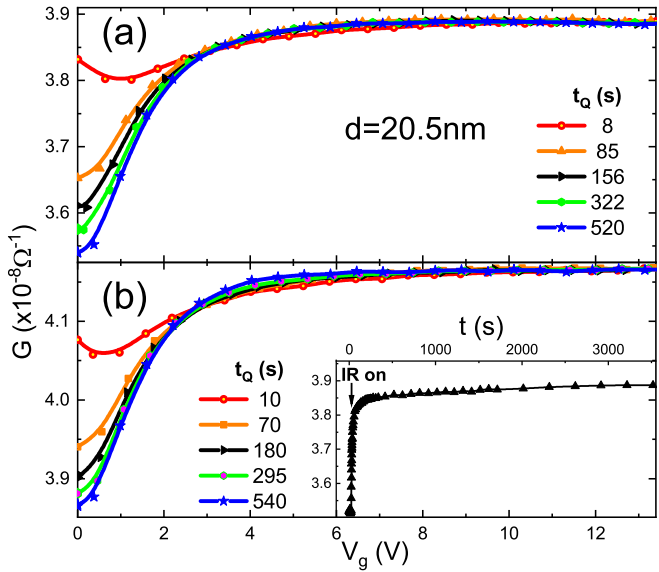}%
\caption{The same as in Fig.4a and Fig.4b respectively except that data are
based on measurements of a 20.5nm In$_{\text{x}}$O sample with a higher degree
of disorder, k$_{\text{F}}\ell$=0.052.}%
\end{figure}

In both samples, the weak-IR decreased the magnitude of the MD, but the
G(V$_{\text{g}}$,t$_{\text{Q}}$) plots converged to a common curve suggesting
that the reduced range of influence associated with $\partial$V$_{\text{IR}}%
$\textperiodcentered f$_{\text{ph}}$ is still larger than the thickness of
these samples in accord with the above estimate for r$_{\text{inf}}$. At the
same time, the effect of the weak IR illumination on the 20nm films, as shown
in Fig.5, is qualitatively similar to the thick film (Fig.2). Quantitatively
however, the reduction of the MD magnitude by the weak-IR is by a factor of
$\approx$2 as compared with $\approx$5.4 found in the 85nm film. The larger
effect in the latter case is another aspect of the cause for the
non-converging G(V$_{\text{g}}$,t$_{\text{Q}}$) plots: Once r$_{\text{inf}}$
(or L$_{\phi}$) becomes smaller than the film thickness, the $\delta$G due to
$\delta$V$_{\text{g}}$ is shunted by the conductance of the unaffected top
layer, thus further reducing the value of $\delta$G/G relative to the
thin-film case.%
\begin{figure}[ptb]%
\centering
\includegraphics[
height=3.3849in,
width=3.4402in
]%
{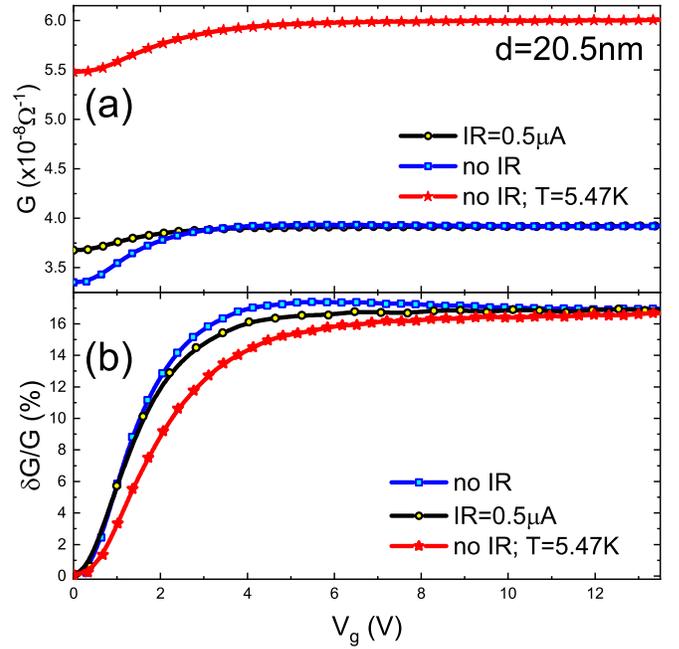}%
\caption{The G(V$_{\text{g}}$) of the 20.5nm sample taken after several hours
of relaxation at V$_{\text{g}}$=0V. Curves are labeled by the measurement
conditions. (a) The raw G(V$_{\text{g}}$) curves. (b) Comparing the MD shapes
(derived from the raw data) after normalizing them by a constant factor f to
match the linear-response curve at V$_{\text{g}}$=0V and V$_{\text{g}}$=13V.
The multiplying factors f for the $\delta$G/G where: 1.0, 2.04, and 1.74 for
the linear-response, IR, and the 5.47K plots respectively.}%
\end{figure}

To conclude, the set of empirical evidence described in this work support the
conjecture that the range of the disturbance caused by varying the
gate-voltage may be limited by introducing nonequilibrium phonons. The
associated high frequency of the ensuing potential modulation presumably
violates the condition of adiabaticity. It is noteworthy that, as long as the
modulation amplitude is weak enough, then up to THz frequencies, this
limitation still allows for nonlocal influence on a significantly large scale
of $\gtrsim$50nm. This should make it feasible to observe quantum effects such
as the orthogonality catastrophe \cite{1}, which is anticipated to be
significant in the Anderson insulating phase \cite{13}. It would also be
relevant for observing modifications of ac conductivity in disordered solids
\cite{14} over a wide range of frequencies. The pre-requirement for these and
other long-scale quantum-effects is that \textit{phase-coherence }must extend
throughout the system to start with.\textit{ }This caveat is rarely mentioned
explicitly in theoretical papers but it is far from a trivial issue in
reality. Unfortunately, little attention has been given to develop tools for
assessing the spatial range of coherence in strongly-disordered insulators,
and very few experimental studies were dedicated to look for quantum effects.
Anisotropic magnetoconductance has been reported in 2D films \cite{15} and
quasi-1D wires \cite{16} of Anderson insulating In$_{\text{2}}$O$_{\text{3-x}%
}$ samples consistent with quantum-interference mechanism, and by inference,
demonstrating quantum coherence on scales that exceed the localization length
of the system. On the other hand, there are systems where these
quantum-effects, while clearly observable in the diffusive regime, vanished or
became overwhelmed by another mechanism once the system crossed over to the
insulating side \cite{17}. It appears that in some materials a dephasing
mechanism is turned on in their localized phase, possibly related to the
appearance of local magnetic-moments associated with singly-occupied sites
\cite{18}. Unless compensated by some mechanism, these are a potential source
of dephasing \cite{19}. This fundamental issue is relevant for all aspects of
quantum transport and it deserves serious experimental and theoretical elucidation.

\begin{acknowledgments}
Discussions with Oded Agam, and Ady Vaknin are gratefully acknowledged. This
research has been supported by a grant No 1030/16 administered by the Israel
Academy for Sciences and Humanities.
\end{acknowledgments}

\end{document}